\documentclass[prd,reprint,showpacs,showkeys]{revtex4}
\usepackage{amsfonts}
\usepackage{amssymb}
\usepackage{amsmath}
\usepackage{graphicx}
\usepackage[font={footnotesize,it}]{caption}

\begin{document}

\title{Global Monopole metric in 2+1-dimensions}
\author{S. Habib Mazharimousavi}
\email{habib.mazhari@emu.edu.tr}
\author{M. Halilsoy}
\email{mustafa.halilsoy@emu.edu.tr}
\affiliation{Department of Physics, Eastern Mediterranean University, Gazima\~{g}usa,
Turkey. }
\date{\today }

\begin{abstract}
In order to obtain the geometry of a global monopole without cosmological
constant and electric charge in $2+1-$ dimensions we make use of the broken $%
O(2)$ symmetry. In the absence of exact solution we determine the series
solutions for both the metric and monopole functions in a consistent manner
that satisfy all equations in appropriate powers. The new expansion elements
are of the form $\frac{1}{r^{n}}\left( \ln r\right) ^{m},$ for the radial
distance $r$ and positive integers $m$ and $n$ constrained by $m\leq n$. To
the lowest order of expansion we find that in analogy with the negative
cosmological constant the geometry of the global monopole acts repulsively,
i.e., in the absence of a cosmological constant the global monopole plays at
large distances the role of a negative cosmological constant.
\end{abstract}

\pacs{}
\keywords{2+1-dimensions; Global monopole; Integral equations;}
\maketitle

\section{Introduction}

Considerable attention received by the lower dimensional field theory and
general relativity during recent decades provides the main motivation for
the present study. It is not only that it constitutes a\ test bed for higher
dimensions but $2+1-$dimensions can also be considered as a brane in $3+1-$%
dimensions. Historically the idea was popularized first by the BTZ black
hole solution \cite{1,2} which was sourced by a cosmological constant. The
charged BTZ black hole was introduced in \cite{3} (two review papers on $%
2+1- $dimensional gravity are \cite{4,5}). From this token the analogue of
Barriola and Vilenkin's global monopole spacetime \cite%
{6,7,8,9,10,11,12,13,14,15,16,17} in $3+1-$dimensions is aimed to be
constructed in $2+1-$dimensions without a cosmological constant. That is, as
the absence of gravitational degrees of freedom was filled by a cosmological
constant in the BTZ black hole spacetime in the present study we propose the
global monopole for the same task. The global monopole is localized in a
core whose effects vanish asymptotically which bends light and yields a
deficit angle in analogy with the $3+1-$dimensional case. Since $2+1-$%
dimension serves as a brane to higher dimensions there are benefits in
studying the lower dimensions. We recall that the global monopole in $3+1-$%
dimensions has the symmetry group $O\left( 3\right) $ to be broken
spontaneously to $U\left( 1\right) $ \cite{18,19,20}. The similar role is
played in the $2+1-$dimensional case by the abelian group $O\left( 2\right)
. $ Instead of a triplet of scalar fields we have now a doublet of scalar
fields $\phi ^{a}=\eta f\left( r\right) \frac{x^{a}}{r}$ (for $a=1,2$) with $%
\eta =$monopole charge constant, $f\left( r\right) $ a radial function to be
determined and $\left( x^{a}\right) ^{2}=r^{2}$ \cite{21,22}. In addition to
the Einstein-global monopole equations through the energy-momentum tensor of
the monopole we have the highly non-linear field equation satisfied by $%
f\left( r\right) ,$ which makes the crux of the problem. The differential
equation satisfied by $f\left( r\right) $ can't be solved exactly, however,
far from the monopole's core we can set $f\left( r\right) =1$. In this paper
we wish to obtain a perturbative solution for $f\left( r\right) $ to
arbitrarily higher orders, yet for technical reasons we terminate our series
at a certain order. By investigating the solution at an arbitrary order we
observe that it admits black holes whose properties are dependent on the
order of perturbation. Recalling the case of BTZ black hole one had to
handle the Einstein-cosmological constant problem which gave the metric
function in a closed form. This is the outcome of a space-time filling
cosmological constant in which one has no other equations to solve. The
presence of a global monopole on the other hand breaks the existing symmetry
and the radial dependent function $f\left( r\right) $ (and therefore $\phi
^{a}\left( r\right) $), adds considerable complexity that didn't exist in
the case of a cosmological constant. The black hole formed from a global
monopole depends strongly on the monopole and its properties. Let us note
from the field theory that in $2+1-$dimensional flat spacetime the
equivalent problem of symmetry breaking yields regular, stable, solitary
structures known as vortices. Such vortex solutions were initiated first by
Nielsen-Olesen \cite{NO} which represent topological structures with finite
energy. Our problem also amounts to determine such irrotational vortex
solutions with the spontaneously broken $U\left( 1\right) $ group but all
taking place in a curved spacetime. An important point that makes our
solution physically significant is the absence of singularity beyond the
core of the monopole. This is a physical requirement since no vortex
solution in any physical theory is allowed to admit singularity. Our
solution is shown numerically to pass this test as revealed by Fig. 4.

For convenience we introduce a new variable $z=\frac{\delta }{r}$ (where $%
\delta =$constant to be defined below) such that the asymptotic region is
mapped to $z\rightarrow 0.$ We expand our metric and monopole functions as $%
f\left( z\right) =\sum_{n=0}^{\infty }\sum_{m=0}^{n}a_{nm}z^{n}\ln ^{m}z$,
where the coefficients $a_{nm}$ are to be determined. By this method we
observe that consistent series solution to the problem of global monopole
can be found to arbitrary powers of $z$. Let us add that any single series
of the form $\sum_{n}a_{n}z^{n}$, $\sum_{n}b_{n}\left( \ln z\right) ^{n}$ or 
$\sum_{n}c_{n}\left( z\ln z\right) ^{n}$ with appropriate constant
coefficients fails to solve the perturbative global monopole problem. The
failure is in the sense that one obtains contradictory expressions in
simultaneous treatment of Einstein-global monopole and the monopole
equations. For this reason we appeal to the novel form of the double series $%
\sum_{n=0}^{\infty }\sum_{m=0}^{n}a_{nm}z^{n}\ln ^{m}z$ which helps to solve
the problem consistently to arbitrarily higher orders of $m$ and $n$.

Ultimately the solution obtained as black hole will have its physical /
thermodynamical properties depending on the integers $m$ and $n$. The
picture turns out to be entirely different from the case of BTZ, which has
its own merits. To our knowledge the global monopole problem in $2+1-$%
dimensions has not been considered elsewhere. We remark that even in a flat
spacetime the monopole equation remains challenging \cite{23}.

\section{Global monopole in $2+1-$dimensions}

We start with the general form of static, circularly symmetric line element
in $2+1-$dimensions given by%
\begin{equation}
ds^{2}=-A\left( r\right) dt^{2}+\frac{1}{B(r)}dr^{2}+r^{2}d\theta ^{2}
\end{equation}%
in which $A\left( r\right) $ and $B\left( r\right) $ are functions only of $%
r $. Now, we introduce the action consisting of a doublet of real scalar
fields of the form 
\begin{equation}
S=\int d^{3}x\sqrt{-g}\left( \frac{R}{2\kappa }+L^{field}\right)
\end{equation}%
in which%
\begin{equation}
L^{field}=-\frac{1}{2}\partial _{\mu }\phi ^{a}\partial ^{\mu }\phi ^{a}-%
\frac{1}{4}\lambda \left( \phi ^{a}\phi ^{a}-\eta ^{2}\right) ^{2}.
\end{equation}%
Here $a=1,2,$ $R$ is the Ricci scalar, $\lambda $ is a coupling constant, $%
\eta $ is the symmetry-breaking scale parameter and%
\begin{equation}
\phi ^{a}=\eta f\left( r\right) \frac{x^{a}}{r},
\end{equation}%
for $x^{1}=r\cos \theta $ and $x^{2}=r\sin \theta .$ To find the field
equation for $f\left( r\right) $ we express the field Lagrangian in terms of 
$f\left( r\right) $ only, i.e., 
\begin{equation}
L^{field}=-\frac{\eta ^{2}B}{2}f^{\prime 2}-\frac{\eta ^{2}}{2r^{2}}f^{2}-%
\frac{1}{4}\lambda \eta ^{4}\left( f^{2}-1\right) ^{2}.
\end{equation}%
Now, variation of the action with respect to $f$ yields%
\begin{equation}
f^{\prime \prime }+\left( \frac{1}{r}+\frac{1}{2AB}\left( AB\right) ^{\prime
}\right) f^{\prime }-\left( \frac{1}{r^{2}}+\lambda \eta ^{2}\left(
f^{2}-1\right) \right) \frac{f}{B}=0
\end{equation}%
in which a prime stands for the derivative with respect to $r.$ Let us note
that even for a flat spacetime background, i.e., $A=B=1,$ Eq. (6) lacks an
exact solution. For this reason in a separate study \cite{23} we have
discussed the possible perturbative solution as a simulation of a Morse type
potential. Variation with respect to $g^{\mu \nu }$ yields the Einstein
equations%
\begin{equation}
G_{\mu }^{\nu }=\kappa T_{\mu }^{\nu }
\end{equation}%
in which 
\begin{equation}
T_{\mu }^{\nu }=\frac{1}{2}\left( \partial _{\mu }\phi ^{a}\partial ^{\nu
}\phi ^{a}-\frac{1}{2}\partial _{\rho }\phi ^{a}\partial ^{\rho }\phi
^{a}\delta _{\mu }^{\nu }\right) -\frac{1}{8}\lambda \left( \phi ^{a}\phi
^{a}-\eta ^{2}\right) ^{2}\delta _{\mu }^{\nu }.
\end{equation}%
An explicit calculation gives%
\begin{equation}
T_{t}^{t}=-\frac{\eta ^{2}}{4}\left( Bf^{\prime 2}+\frac{1}{r^{2}}f^{2}+%
\frac{\lambda }{2}\eta ^{2}\left( f^{2}-1\right) ^{2}\right) ,
\end{equation}%
\begin{equation}
T_{r}^{r}=\frac{\eta ^{2}}{4}\left( Bf^{\prime 2}-\frac{1}{r^{2}}f^{2}-\frac{%
\lambda }{2}\eta ^{2}\left( f^{2}-1\right) ^{2}\right)
\end{equation}%
and%
\begin{equation}
T_{\theta }^{\theta }=-\frac{\eta ^{2}}{4}\left( Bf^{\prime 2}-\frac{1}{r^{2}%
}f^{2}+\frac{\lambda }{2}\eta ^{2}\left( f^{2}-1\right) ^{2}\right) .
\end{equation}

In addition, the Einstein tensor's components are given by%
\begin{equation}
G_{t}^{t}=\frac{1}{2}\frac{B^{\prime }}{r},
\end{equation}%
\begin{equation}
G_{r}^{r}=\frac{1}{2}\frac{A^{\prime }B}{rA}
\end{equation}%
and%
\begin{equation}
G_{\theta }^{\theta }=\frac{1}{4}\frac{2A^{\prime \prime }AB-A^{\prime
2}B+A^{\prime }B^{\prime }A}{A^{2}}.
\end{equation}%
As we are interested in a solution at large $r$ we define $z=\frac{\delta }{r%
}$ (with $\delta =\eta \sqrt{\lambda }$ the size of the global monopole with
the unit of length so that $z$ becomes a dimensionless parameter) and
rewrite the field equations in terms of $z.$ Also we set $8\pi G=\kappa =%
\frac{1}{\eta ^{2}}$ while $c=1.$ The Einstein equations are given by (note
that without loss of generality we set $\delta =1$)%
\begin{equation}
-\frac{1}{2}z^{3}B^{\prime }+\frac{1}{4}z^{4}Bf^{\prime 2}+\frac{1}{4}%
z^{2}f^{2}+\frac{1}{8}\left( f^{2}-1\right) ^{2}=0,
\end{equation}%
\begin{equation}
-\frac{1}{2}z^{3}B\frac{A^{\prime }}{A}-\frac{1}{4}z^{4}Bf^{\prime 2}+\frac{1%
}{4}z^{2}f^{2}+\frac{1}{8}\left( f^{2}-1\right) ^{2}=0,
\end{equation}%
\begin{equation}
\frac{1}{2}\frac{Bz^{3}}{A}\left( zA^{\prime \prime }+2A^{\prime }\right) +%
\frac{z^{4}A^{\prime }}{4A^{2}}\left( B^{\prime }A-BA^{\prime }\right) +%
\frac{1}{4}z^{4}Bf^{\prime 2}-\frac{1}{4}z^{2}f^{2}+\frac{1}{8}\left(
f^{2}-1\right) ^{2}=0
\end{equation}%
in which a prime denotes $\frac{d}{dz}$ and finally the field equation for $%
f\left( z\right) $ takes the form%
\begin{equation}
z^{4}f^{\prime \prime }+2z^{3}f^{\prime }+z^{2}f^{\prime }\left( -z+\frac{%
z^{2}A^{\prime }}{2A}+\frac{z^{2}B^{\prime }}{2B}\right) -\frac{\left(
z^{2}+f^{2}-1\right) f}{B}=0.
\end{equation}%
Combination of the first two equations yields%
\begin{equation}
A=B\Delta ^{2}
\end{equation}%
in which 
\begin{equation}
\Delta =C_{0}\exp \left( -\frac{1}{2}\int zf^{\prime 2}dz\right)
\end{equation}%
where we shall set the integration constant $C_{0}=1$ by knowing that it can
be absorbed into the redefinition of time. Using (15) we find 
\begin{equation}
B=\frac{1}{\Delta }\left( C+\int \frac{\Delta \left(
2f^{2}z^{2}+f^{4}-2f^{2}+1\right) }{4z^{3}}dz\right)
\end{equation}%
with $C$ another integration constant. Next, we substitute $A$ and $B$ into
the other two equations which reduce to the same equation given by%
\begin{equation}
\left( 1+f^{4}+2\left( z^{2}-1\right) f^{2}\right) zf^{\prime }+4f\left(
1-z^{2}-f^{2}\right) +\frac{4z^{3}\left( \Omega +C\right) \left( f^{\prime
}+zf^{\prime \prime }\right) }{\Delta }=0
\end{equation}%
where $\Omega =\int \frac{\Delta \left( 2f^{2}z^{2}+f^{4}-2f^{2}+1\right) }{%
4z^{3}}dz.$ 
\begin{figure}[tbp]
\includegraphics[width=70mm,scale=0.7]{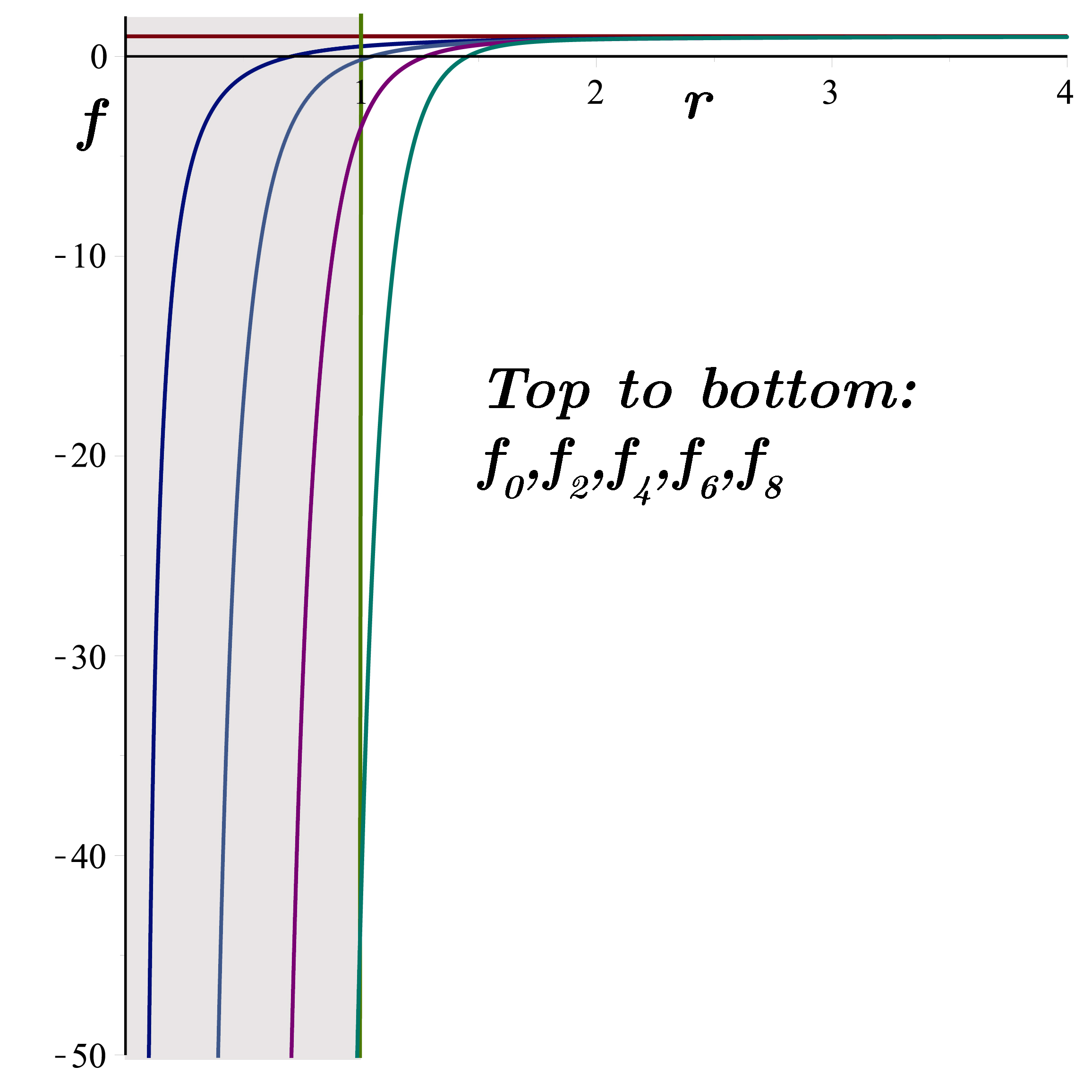} %
\captionsetup{justification=raggedright, singlelinecheck=false}
\caption{A plot of $f$ versus $r$ for $C=0.3.$ The subindex stands for the
power of $z$ in the expression of $f.$ The core of the monopole is shown as
shaded area. }
\end{figure}
\begin{figure}[tbp]
\includegraphics[width=70mm,scale=0.7]{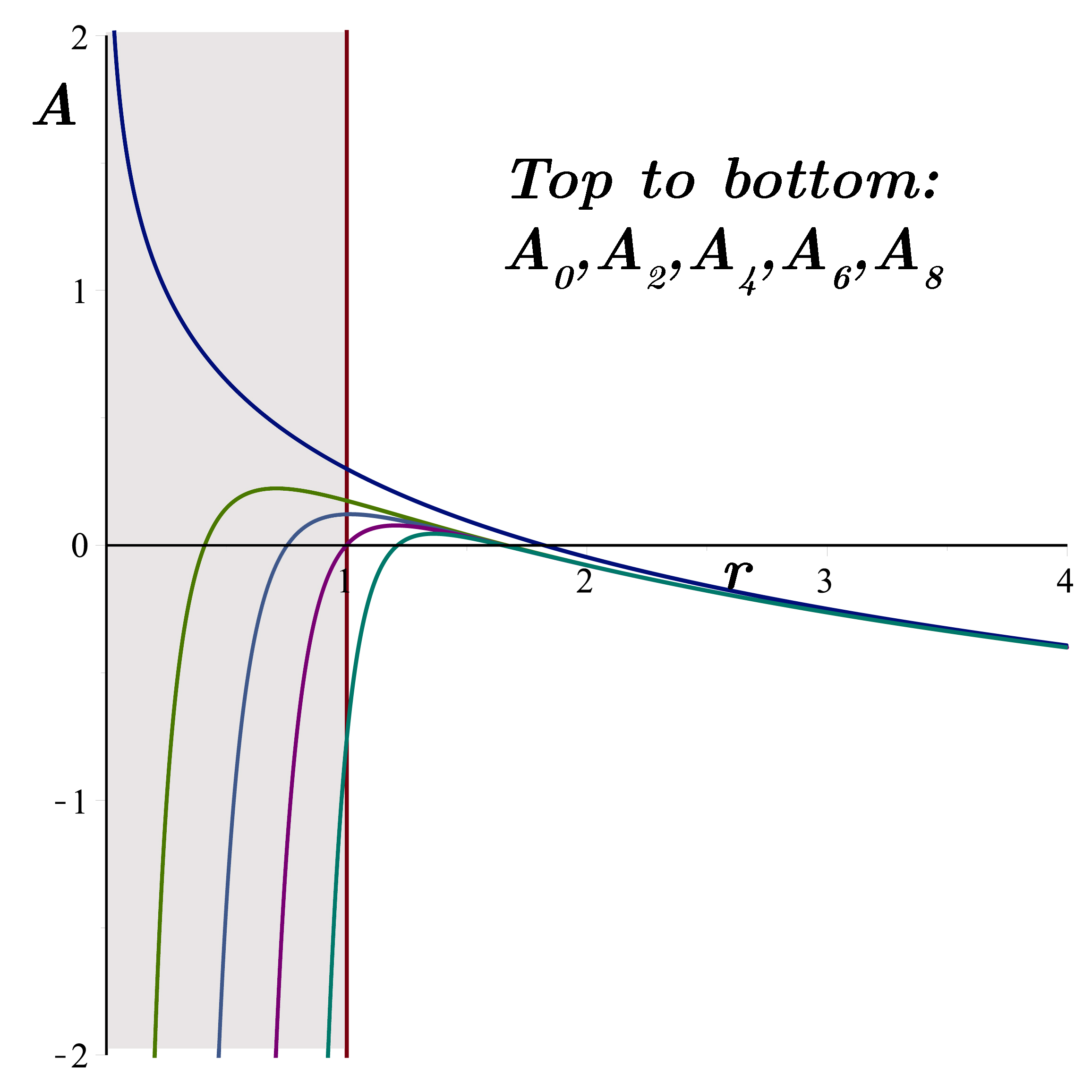} %
\captionsetup{justification=raggedright, singlelinecheck=false}
\caption{A plot of $A$ versus $r$ for $C=0.3.$ The subindex stands for the
power of $z$ in the expression of $A.$ The event horizon and the
cosmological horizon imply the solution is a black hole. The shaded area is
the core of the global monopole.}
\end{figure}
\begin{figure}[tbp]
\includegraphics[width=70mm,scale=0.7]{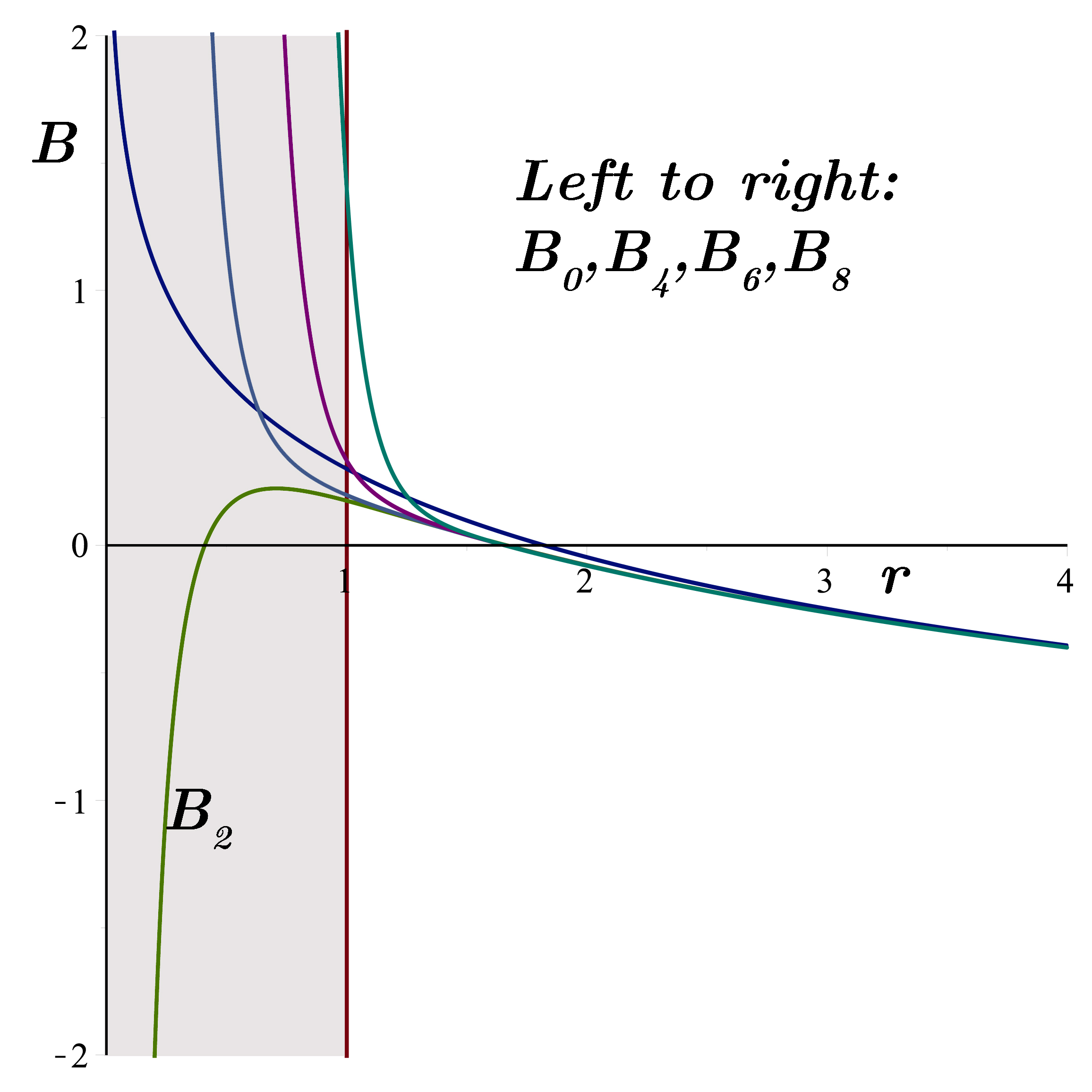} %
\captionsetup{justification=raggedright, singlelinecheck=false}
\caption{A plot of $B$ versus $r$ for $C=0.3.$ The subindex stands for the
order of $z$ in the expression of $B.$ The shaded area is the core of the
global monopole. Let's add that, from $B_{0}$ to $B_{4}$ there are some
fluctuations, but after that the solution becomes stable.}
\end{figure}

\begin{figure}[tbp]
\includegraphics[width=70mm,scale=0.7]{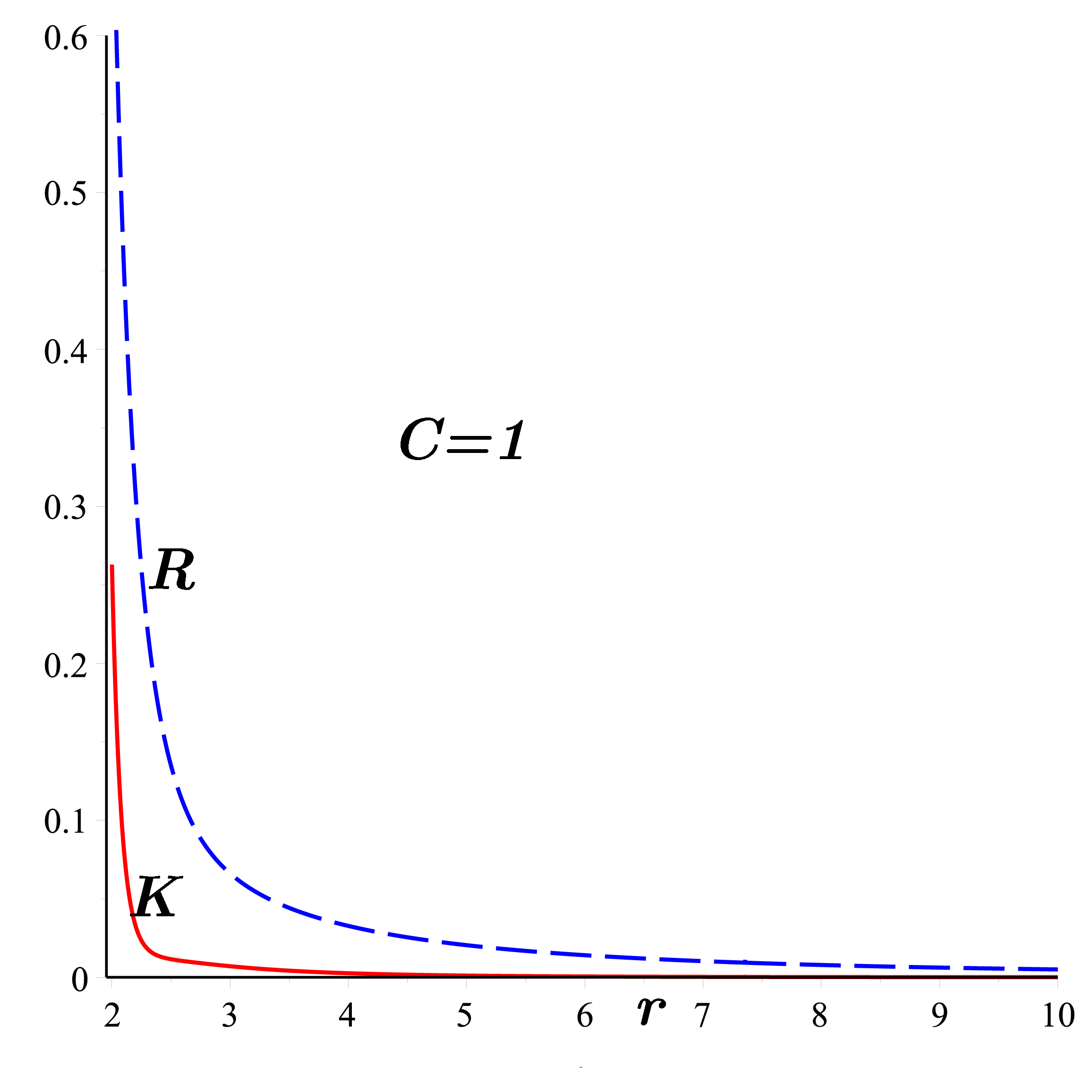} %
\captionsetup{justification=raggedright, singlelinecheck=false}
\caption{Ricci scalar (blue-dash) and Kretschmann scalar (Red-solid) in
terms of $r$ outside ($r>2\protect\delta $) the core of the monopole. To
plot $R$ and $K$ we employed $A$ and $B$ up to the power $8$ given in (25)
and (26). Regularity of our series solution is evidently seen from these
plots.}
\end{figure}
As we look for a solution for $f$ which asymptotically approaches to the one
with conditions, $f\left( 0\right) =1$ and $f^{\prime }\left( 0\right) =0,$
we set a series solution of the form%
\begin{equation}
f=\sum_{n=0}^{\infty }\sum_{m=0}^{n}a_{nm}z^{n}\ln ^{m}z
\end{equation}%
in which $a_{nm}$ are constant coefficients to be found. Our results reveal
that this ansatz works to any higher order one wishes and the following is
its explicit form up to order $z^{8},$%
\begin{multline}
f=1-\frac{z^{2}}{2}-z^{4}\left( \frac{3}{8}+C+\frac{1}{2}\ln z\right)
-z^{6}\left( \frac{9}{16}+7C+8C^{2}+\left( \frac{7}{2}+8C\right) \ln z+2\ln
^{2}z\right) - \\
z^{8}\left( \frac{187}{128}+52C+\frac{391C^{2}}{2}+144C^{3}+\left( 26+\frac{%
391}{2}C+216C^{2}\right) \ln z+\left( 108C+\frac{391}{8}\right) \ln
^{2}z+18\ln ^{3}z\right) +\mathcal{O}\left( z^{10}\right) .
\end{multline}%
Upon this result we also find the series solutions for $A$ and $B$ as follow%
\begin{multline}
A=C+\frac{1}{2}\ln z-\frac{z^{2}}{8}-z^{4}\left( \frac{C}{8}+\frac{1}{64}+%
\frac{1}{16}\ln z\right) +z^{6}\left( \frac{7}{864}-\frac{5C}{18}-\frac{C^{2}%
}{2}-\left( \frac{C}{2}+\frac{5}{36}\right) \ln z-\frac{1}{8}\ln
^{2}z\right) + \\
z^{8}\left( \frac{253}{9216}-\frac{127C}{192}-\frac{81C^{2}}{16}%
-5C^{3}-\left( \frac{127}{384}+\frac{81C}{16}+\frac{15C^{2}}{2}\right) \ln
z-\left( \frac{81}{64}+\frac{15C}{4}\right) \ln ^{2}z-\frac{5}{8}\ln
^{3}z\right) +\mathcal{O}\left( z^{10}\right) 
\end{multline}%
and%
\begin{multline}
B=C+\frac{1}{2}\ln z-\frac{z^{2}}{8}+z^{4}\left( \frac{C}{8}-\frac{1}{64}+%
\frac{1}{16}\ln z\right) +z^{6}\left( -\frac{5}{216}-\frac{5C}{18}-\frac{%
5C^{2}}{6}+\left( \frac{5C}{6}+\frac{5}{36}\right) \ln z+\frac{5}{24}\ln
^{2}z\right) + \\
z^{8}\left( -\frac{47}{1024}+\frac{19C}{32}+\frac{123C^{2}}{16}%
+9C^{3}+\left( \frac{19}{64}+\frac{123C}{16}+\frac{27C^{2}}{2}\right) \ln
z+\left( \frac{123}{64}+\frac{27C}{4}\right) \ln ^{2}z+\frac{9}{8}\ln
^{3}z\right) +\mathcal{O}\left( z^{10}\right) .
\end{multline}%
%
%
\begin{figure}[tbp]
\includegraphics[width=70mm,scale=0.7]{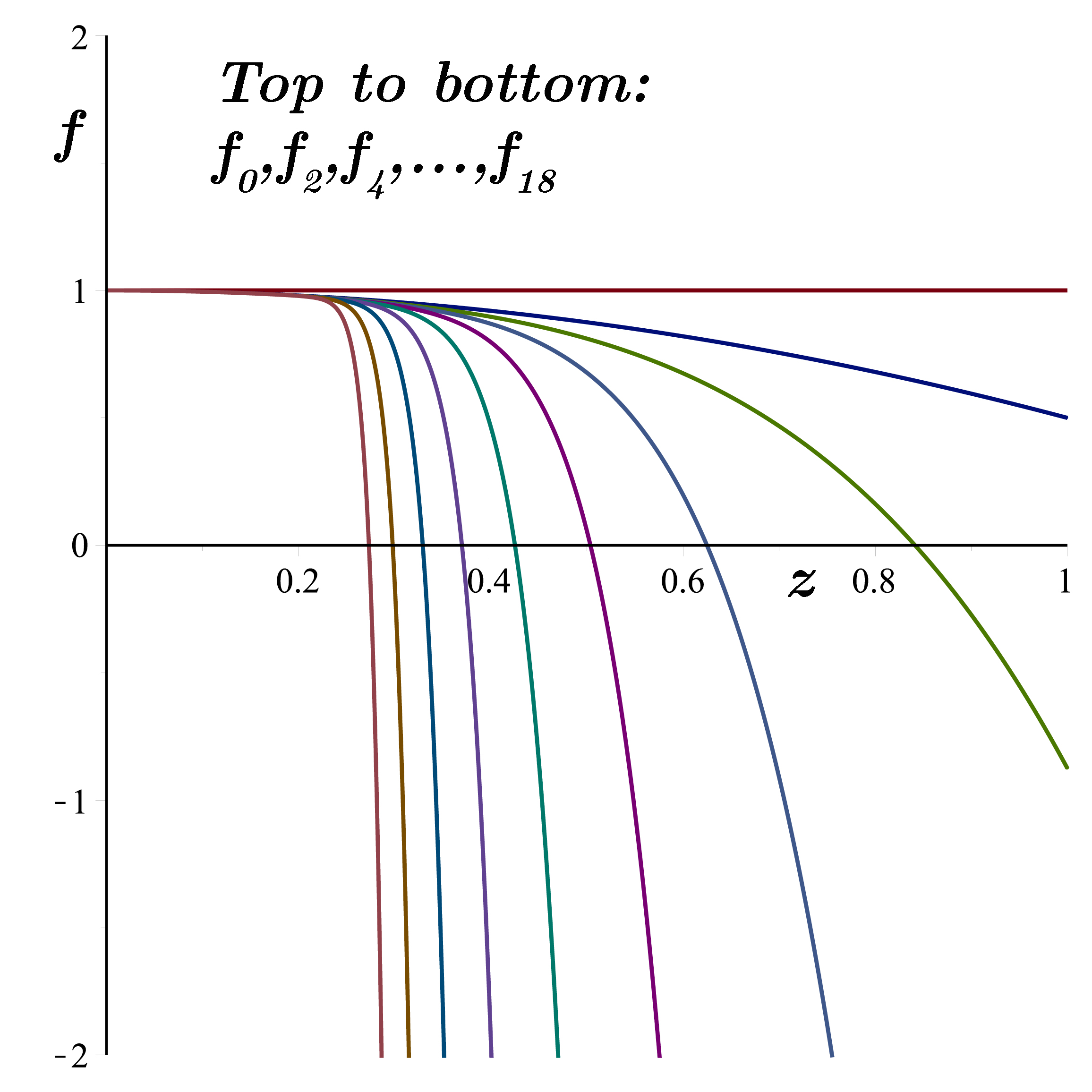} %
\captionsetup{justification=raggedright, singlelinecheck=false}
\caption{Plot of $f\left( z\right) $ with respect to $z$ with C=1. The
subindices imply the maximum power of $z$. It is seen that the
series converges slowly but for sure at near zero all of them coincide.}
\end{figure}
\begin{figure}[tbp]
\includegraphics[width=70mm,scale=0.7]{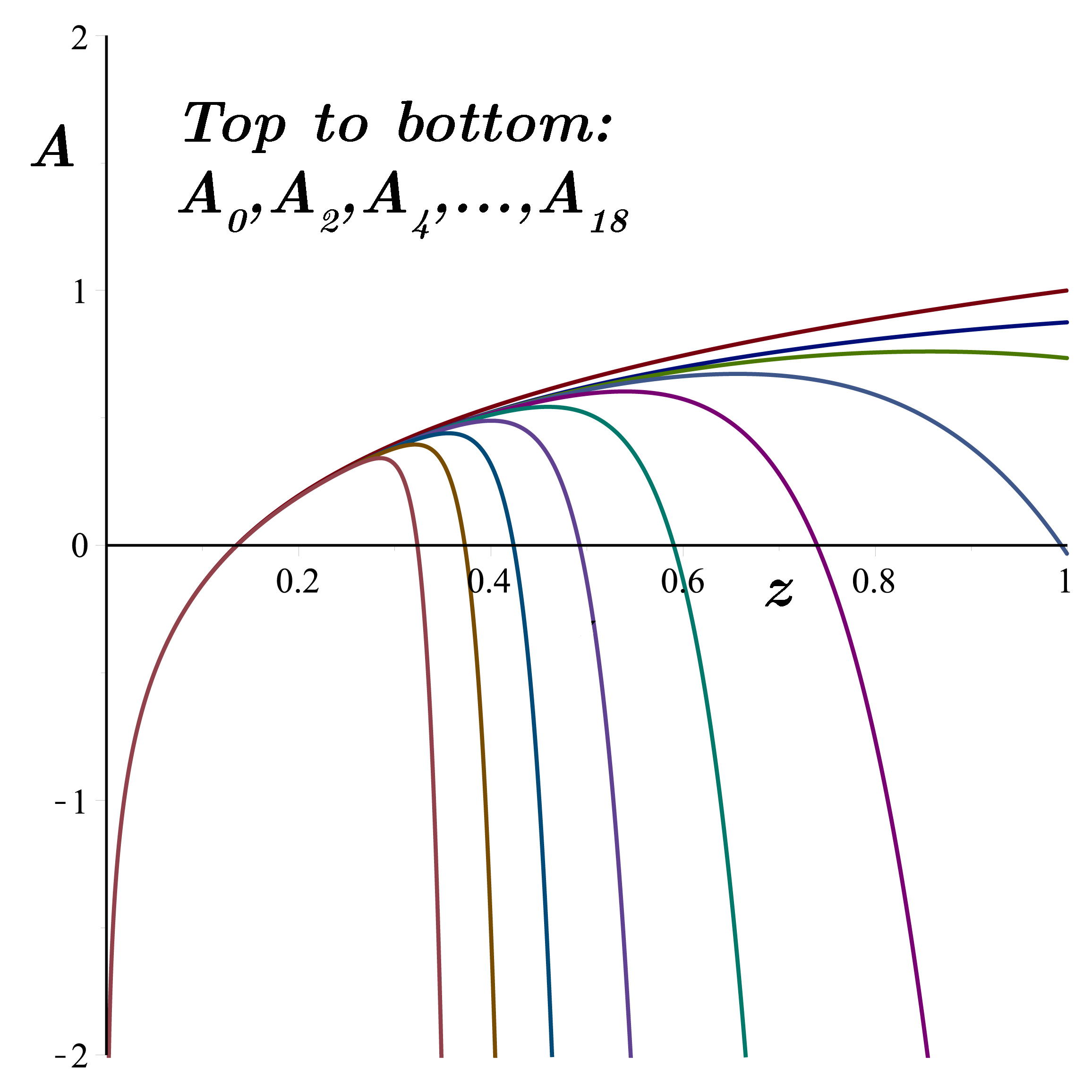} %
\captionsetup{justification=raggedright, singlelinecheck=false}
\caption{Plot of $A\left( z\right) $ with respect to $z=0$ with C=1. The
subindices imply the maximum power of $z$. Similar to $f\left( z\right) $
one observes that the series is converging slowly but at near zero all of
them coincide.}
\end{figure}
\begin{figure}[tbp]
\includegraphics[width=70mm,scale=0.7]{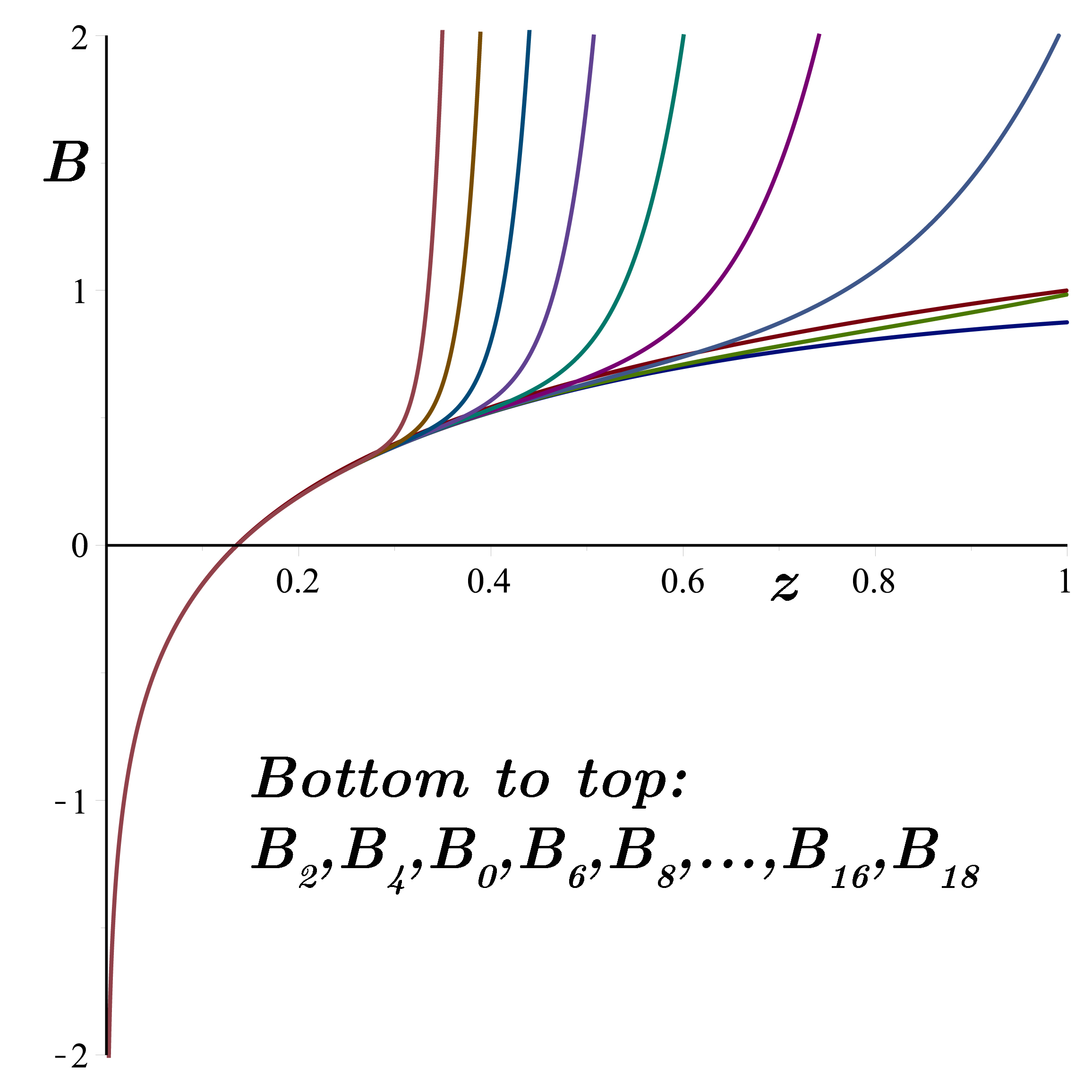} %
\captionsetup{justification=raggedright, singlelinecheck=false}
\caption{Plot of $B\left( z\right) $ with respect to $z=0$ with C=1. The
subindices imply the maximum power of $z$. Except for the first three terms
i.e., $B_{0,}B_{2}$ and $B_{4}$ the rest of the terms appear to be
converging. Near $z=0$ all curves coincide.}
\end{figure}

We note that the solution is valid for outside the global monopole, i.e., $%
r>\delta $ in (24)-(26), with $0<z<1.$ Let us comment that in contrast to
the effect of mass, global monopole creates a repulsive force. This can
easily be seen in the lowest order where $A\left( r\right) =C-\frac{1}{2}\ln 
\frac{r}{\delta }-\frac{\delta ^{2}}{8r^{2}},$ which corresponds to the
Newtonian force proportional to $\frac{1}{4r}>0$. In Figs. 1-3 we plot $f,$ $%
A$ and $B$ in terms of $r=\frac{\delta }{z}$ for $C=1.$ The subindex stands
for the power of $z,$i.e., $f_{0}=1,$ $f_{2}=1-\frac{z^{2}}{2},$ $f_{4}=1-%
\frac{z^{2}}{2}-z^{4}\left( \frac{3}{8}+C+\frac{1}{2}\ln z\right) $ and so
on. As one can see inclusion of higher order of $z$ does not change the
behavior of the solutions for $r\gg \delta .$ Also Fig.s 2 and 3 reveals the
black hole feature of the solution that in addition to an event horizon
there exists a cosmological horizon at large $r$ which is independent of the
number of terms we retain. In Fig. 4 we plot the Ricci and Kretchmann
scalars of the solution for large $r$ i.e., $r>2\delta .$

In Fig. 5-7 we plot $f\left( z\right) ,$ $A\left( z\right) $ and $B\left(
z\right) $ in terms of $z$ up to $z^{18}.$ In all of these figures the curves
converge slowly but at small $z$ which means large $r$ all curves coincide.
We must also add that in the problem of global monopole we are looking for a
solution at radiation region. Therefore for our purpose the behavior of the
solutions at near $z=0$ is sufficient.

\subsection{Solution near the origin}

In this section we give a series solution to the main field equations when $%
\frac{\delta }{r}>1.$ As in the previous section, first we combine the $tt$
and $rr$ components of the Einstein equation to get (up to a constant
coefficient which can be absorbed in time)%
\begin{equation}
A\left( r\right) =B\left( r\right) e^{\int r\left( f^{\prime }\right)
^{2}dr}.
\end{equation}%
Putting this into the $tt$ component of the Einstein equations yields%
\begin{equation}
B\left( r\right) =e^{-\frac{1}{2}\int r\left( f^{\prime }\right)
^{2}dr}\left( C-\frac{1}{4}\int \frac{\left(
2f^{2}+r^{2}f^{4}-2r^{2}f^{2}+r^{2}\right) e^{\frac{1}{2}\int r\left(
f^{\prime }\right) ^{2}dr}}{r}dr\right)
\end{equation}%
in which $C$ is an integration constant. By setting $C=1$ we found a regular
series solution for the field $f$ and the metric functions $A$ and $B$ as
follows 
\begin{equation}
f\left( r\right) =ar+a\frac{1}{8}\left( a^{2}-\frac{5}{8}\right) r^{3}+a%
\frac{5}{128}\left( a^{4}-\frac{1}{6}a^{2}-\frac{7}{96}\right) r^{5}+a%
\mathcal{O}\left( r^{7}\right) ,
\end{equation}%
\begin{equation}
B\left( r\right) =1-\frac{1}{2}\left( a^{2}+\frac{1}{4}\right) r^{2}-\frac{%
a^{2}}{16}\left( a^{2}-\frac{7}{2}\right) r^{4}-\frac{a^{2}}{64}\left( a^{4}+%
\frac{43}{24}a^{2}+\frac{165}{128}\right) r^{6}+a^{2}\mathcal{O}\left(
r^{8}\right)
\end{equation}%
and%
\begin{equation}
A\left( r\right) =1-\frac{1}{8}r^{2}+\frac{5}{128}a^{2}r^{4}+\frac{5}{2304}%
a^{2}\left( a^{2}-\frac{1}{2}\right) r^{6}+a^{2}\mathcal{O}\left(
r^{8}\right)
\end{equation}%
in which $a$ is an undetermined nonzero constant and $\delta =1$. Let us
note that for $a=0$ we have $f=0,$ i.e., no monopole case with action%
\begin{equation}
S=\frac{1}{2\kappa }\int d^{3}x\sqrt{-g}\left( R-\frac{1}{2}\right)
\end{equation}%
and $A=B=1-\frac{r^{2}}{8},$ which corresponds to the particular BTZ metric 
\cite{BTZ}. Near the origin we have explicitly a flat spacetime, as it
should be.

\section{Conclusion}

We obtained the metric of a global monopole in $2+1-$dimensions which is in
the absence of cosmological constant as the analogue of the
Barriola-Vilenkin's monopole in $3+1-$dimensions. Inclusion of electric
charge becomes possible with the addition of a Maxwell source and replacing
Einstein equations with the Einstein-Maxwell equations. Outside the core ($%
r>\delta $) we have an asymptotic solution that admits black hole with the
hair of the monopole. Formation of global monopoles are attributed to the
topological remnants as a result of spontaneously broken $O(2)$ symmetry.
Exact analytical solution for the monopole function $f\left( r\right) $
doesn't exist so that in this regard it emerges as tough as in the case of $%
3+1-$dimensions. We recall that the global monopole function $\phi
^{a}\left( r\right) $ (or $f\left( r\right) $) in the Schwarzschild geometry
could also be treated as an expansion in powers of $\frac{1}{r}$ \cite{6}.
We are satisfied with the novel series solution of the form (23) with $z=%
\frac{\delta }{r}$. This is in marked distinction from the series expansion
of the $3+1-$dimensional monopole in which single index (say $n$) expansion
of the series solved the problem asymptotically. In other words, in the $2+1-
$dimensional monopole problem series expansion of the form $\sum a_{n}\frac{1%
}{r^{n}},$ ($a_{n}=$const.) gives inconsistency in the overall problem.
Necessarily therefore both the powers of $r$ and $\ln r$ coexist in the
expansions and our solution beyond the case $(r>\delta )$ is free of
singularity. By this approach the metric functions $A\left( r\right) $, $%
B\left( r\right) $ can be determined in a consistent manner to arbitrarily
higher-orders. We recall that in all studies of global monopoles only the
asymptotic ($r\rightarrow \infty $) behavior has been considered.
Unfortunately we were unable to find a correlation among terms therefore we
can't give the most general term of the series in a closed form.
Technically, however, one can find up to any higher order of $z$ one wishes.
We must add also that the choice of integration constant i.e., $C$ plays a
crucial role. The overall effect of the monopole turns out, at the lowest
order to be repulsive, i.e., as in the positive cosmological constant case
that opposes gravity. Let us complete the paper by proposing that our method
of double-series expansion may find rooms of application in different field
theories that admit vortex solutions; both rotational and irrotational.

\end{document}